\documentclass[a4paper]{article}
\pdfoutput=1
\usepackage[round]{natbib}
\usepackage{amsmath,amsthm,amssymb}
\usepackage{float}
\usepackage{physics}
\usepackage{graphicx}
\usepackage{setspace}
\usepackage{subcaption}
\usepackage{fullpage}
\usepackage{authblk}
\usepackage{lineno}
\usepackage{multicol}
\usepackage{caption}
\usepackage{hyperref}
\hypersetup{
    colorlinks=true,
    citecolor=blue,
}

\begin{document}

\title{Computation of local permeability in gap-graded granular soils}
\author[1]{V. S. Ramakrishna Annapareddy\footnote{Corresponding Author, email:
\href{v.annapareddy@uq.edu.au}{v.annapareddy@uq.edu.au}, ORCiD: 0000-0003-2816-3250}}
\author[1]{Adnan Sufian} 
\author[1]{Thierry Bore} 
\author[2]{Mathieu Bajodek} 
\author[1]{Alexander Scheuermann} 
\affil[1]{School of Civil Engineering, The University of Queensland, Brisbane, Australia}
\affil[2]{LAAS-CNRS, Univ de Toulouse, UPS, Toulouse, France}
\date{}
\setcounter{Maxaffil}{0}
\renewcommand\Affilfont{\itshape\small}

\maketitle

\begin{abstract}
This paper proposes semi-analytical methods to obtain the local permeability for granular soils based on indirect measurements of the local porosity profile in a large coaxial cell permeameter using spatial time-domain reflectometry. 
The porosity profile is used to obtain the local permeability using the modified Kozeny-Carman and Katz-Thompson equations, which incorporated an effective particle diameter that accounted for particle migration within the permeameter. 
The profiles of the local permeability obtained from the proposed methods are compared with experimentally obtained permeability distributions using pressure measurements and flow rate.  
The permeabilities obtained with the proposed methods are comparable with the experimentally obtained permeabilities and are within one order of magnitude deviation, which is an acceptable range for practical applications.
\end{abstract}

\setlength{\parindent}{0em}
\setlength{\parskip}{1em}

\section{Introduction}

The ease with which a fluid flows through the interconnected pore network of the soil is defined as permeability, which is an important material property in many engineering fields including geotechnical engineering, petroleum engineering, hydraulic engineering, environmental engineering and agriculture \citep{Harr1991}. Standard laboratory tests, such as the constant head and falling head tests, are commonly employed to determine soil permeability. However, these laboratory tests only provide a global measure of permeability across the entire sample and fail to capture the effects of soil structure and pore-scale heterogeneity \citep{mishra2020evaluation}. Several studies have demonstrated through numerical investigations the effects of pore-scale properties causing heterogeneities on soil permeability \citep{stewart2006study,van2018computational,liu2019pore,sufian2019ability}. Recently, the microstructure of dual porosity media (aggregated soil, a type of granular soil) has been numerically investigated by \citet{zhang2021fluid}; \citet{zhang2021poroelastic}. However, there are limited physical observations of local permeability in laboratory experimental studies owing to the challenges of the characterisation of pore-scale features. Conventionally, the measurement of local permeability is based on the point measurements of the hydraulic head along the specimen length using pressure transducers (PT) or standpipe piezometers \citep{kenney1985internal,moffat2006large,wan2008assessing}. 
However, this approach is limited to a local observation depending on the position and spacing of the transducers/standpipes making it impossible to capture local heterogeneities within the sample.

Particularly in broadly graded and gap-graded soils, heterogeneities can occur which can significantly influence the overall permeability of a sample despite all efforts to produce it as homogeneously as possible. In addition, gap-graded soils are prone to suffusion phenomena that involve redistribution or loss of fine particles without causing a significant change in soil volume \citep{fannin2014distinct}, which in turn lead to pore-scale changes in the soil structure, and hence, local changes in the porosity and permeability \citep{nguyen2019experimental}. Therefore, physical observations that can capture the spatial and temporal variation of permeability plays an important role in understanding suffusion. Another example where local permeability profile is important is in verifying the effectiveness of the microbial induced calcite precipitation (MICP) method, particularly in determining whether local or uniform improvements have been observed \citep{hataf2020reducing}.

This paper combines indirect measurements of the local porosity with semi-analytical methods to obtain the local permeability profile in granular gap-graded soils. A large coaxial cell permeameter \citep{bittner2019determination,scheuermann2009feasibility,yerman2018integration,yan2021application,annapareddy2021onset} and spatial time-domain reflectometry (STDR) are used to obtain the local porosity \citep{schlaeger2005fast,scheuermann2012determination,robinson2003review,bore2016error,mishra2018dielectric}. STDR enables high resolution physical measurements on the spatial and temporal variation of local permeability, which is not possible in other experimental approaches. This will provide new experimental insights into the pore-scale heterogeneity of gap-graded soils, which has previously only been explored with numerical simulations \citep{sufian2021influence}. The measured local porosity is then used to obtain the local permeability profile using modified Kozeny-Carman (K-C) and Katz-Thompson (K-T) equations. The local permeability profiles obtained from the proposed methods are compared against the conventional approach to obtaining local permeability by considering adjacent pressure transducers, as well as the average permeability across the entire soil layers.

\begin{figure}[ht]
     \centering
     \includegraphics[width=\textwidth]{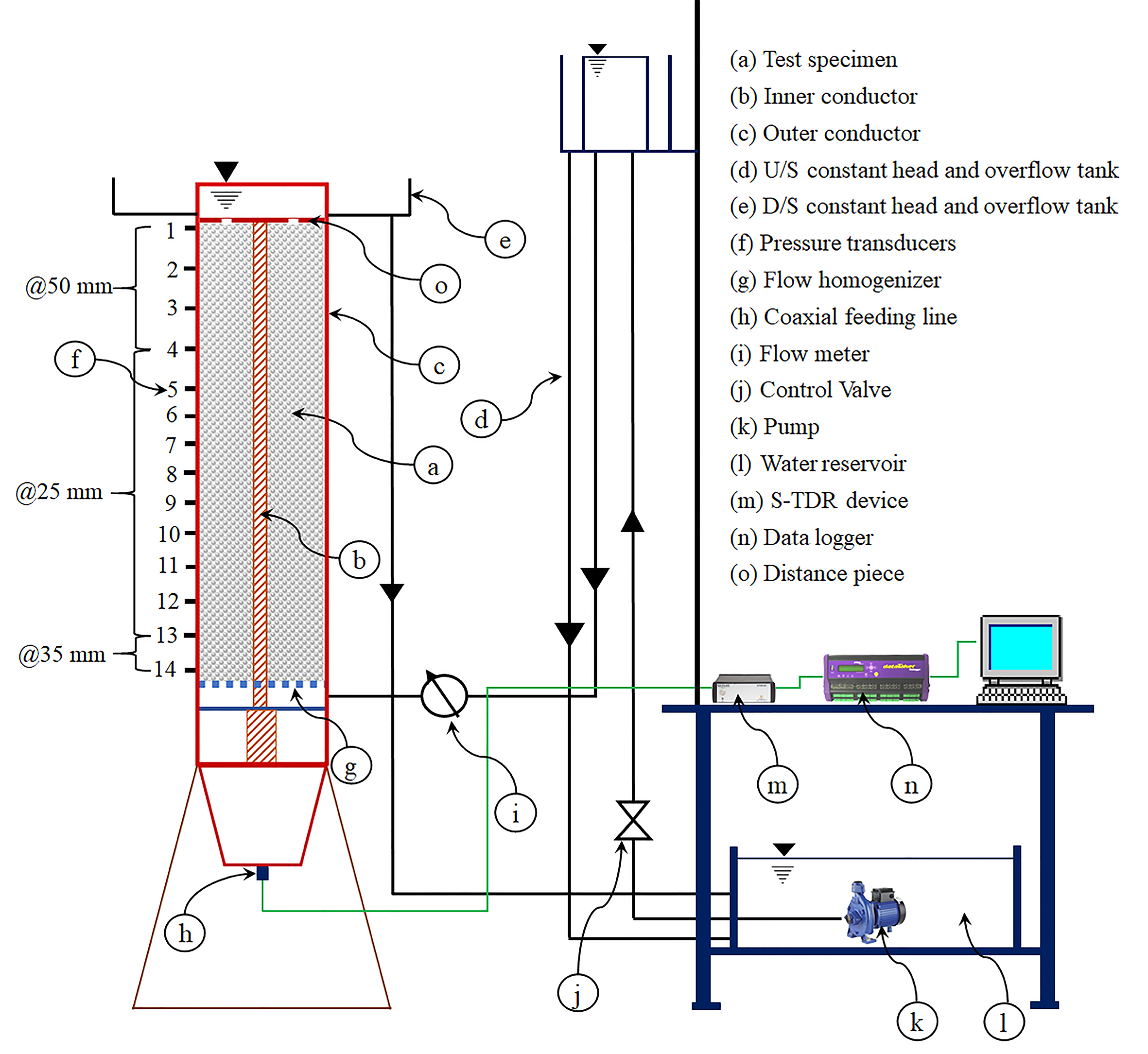}
     \caption{Schematic view of the experimental setup}
     \label{fig:1}
 \end{figure}

\section{Experimental Setup and Test Procedure}

The experimental setup consists of a large coaxial cell permeameter, a hydraulic control system, a spatial time-domain reflectometer device and a data acquisition system [Fig.~\ref{fig:1} and details in \cite{bittner2019determination}]. The copper-built permeameter acts as a coaxial transmission line comprising inner and outer conductors that act as sensors enabling electromagnetic measurements using the spatial time-domain reflectometer device. The outer diameter of the inner conductor is 41.3 mm and the inner diameter of the outer conductor is 151.9 mm. The test specimen is placed in the annulus between the inner and outer conductor. A 4 cm wide observation window is located on the outer walls of the coaxial cell. The hydraulic control system enables vertical upwards flow in a closed loop. The applied hydraulic gradient can be altered by changing the position of the upstream (U/S) constant head tank, whereas, the downstream (D/S) hydraulic head is kept constant through an overflow at the top of the permeameter. Continuous measurements of pore water pressures are obtained from the pressure transducers mounted on the cell wall, while continuous measurements of the flow rate are obtained from a flow-meter.

The sample comprises glass beads of various sizes and the preparation of the sample is graphically shown in Fig.~\ref{fig:2}. The test specimen is placed between the top and bottom filter layers. The bottom filter prevents the loss of fine particles through the flow homogeniser at the base of the sample. As illustrated in Fig.~\ref{fig:2}, the test specimen consists of a mixture layer comprising fine and coarse fractions, which is below a coarse layer comprising entirely of the coarse fraction. The specimen was purposely prepared in this manner to enable STDR to capture the changes in local porosity and permeability caused by the migration of fine particles from the mixture layer to the coarse layer at sufficiently high hydraulic gradients. Fig.~\ref{fig:3} shows the particle size distribution (PSD) of the test specimen and the PSDs of the fine and coarse fractions of the test specimen. The test specimen contains approximately 8.1\% of fines fraction by mass, indicating an under-filled fabric with the finer particles sitting within the voids of the coarse fraction. According to the \citep{kezdi1979} criterion, the test specimen is internally unstable, such that at a critical hydraulic gradient the fine particles within the mixture layer will dislodge and transport fine into the coarse layer. The specimen was saturated by applying a very low gradient, after which the gradient was increased in multiple steps up to a maximum of 1.91. At each increment, the gradient was kept constant for 10 minutes and the total duration of the experiment was approximately 380 minutes. This paper focusses on the ability of STDR to enable the calculation of the local permeability profile, which is demonstrated by considering the conditions at the start of the test when the mixture layer and coarse layer are distinct, and at the end of the test when the finer particles of the mixture layer have migrated into the coarse layer.

\newenvironment{Figure}
  {\par\medskip\noindent\minipage{\linewidth}}
  {\endminipage\par\medskip}

\begin{multicols}{2}

 \begin{Figure}
     \centering
     \captionsetup{type=figure,justification=centering}
     \includegraphics[height=7.5cm]{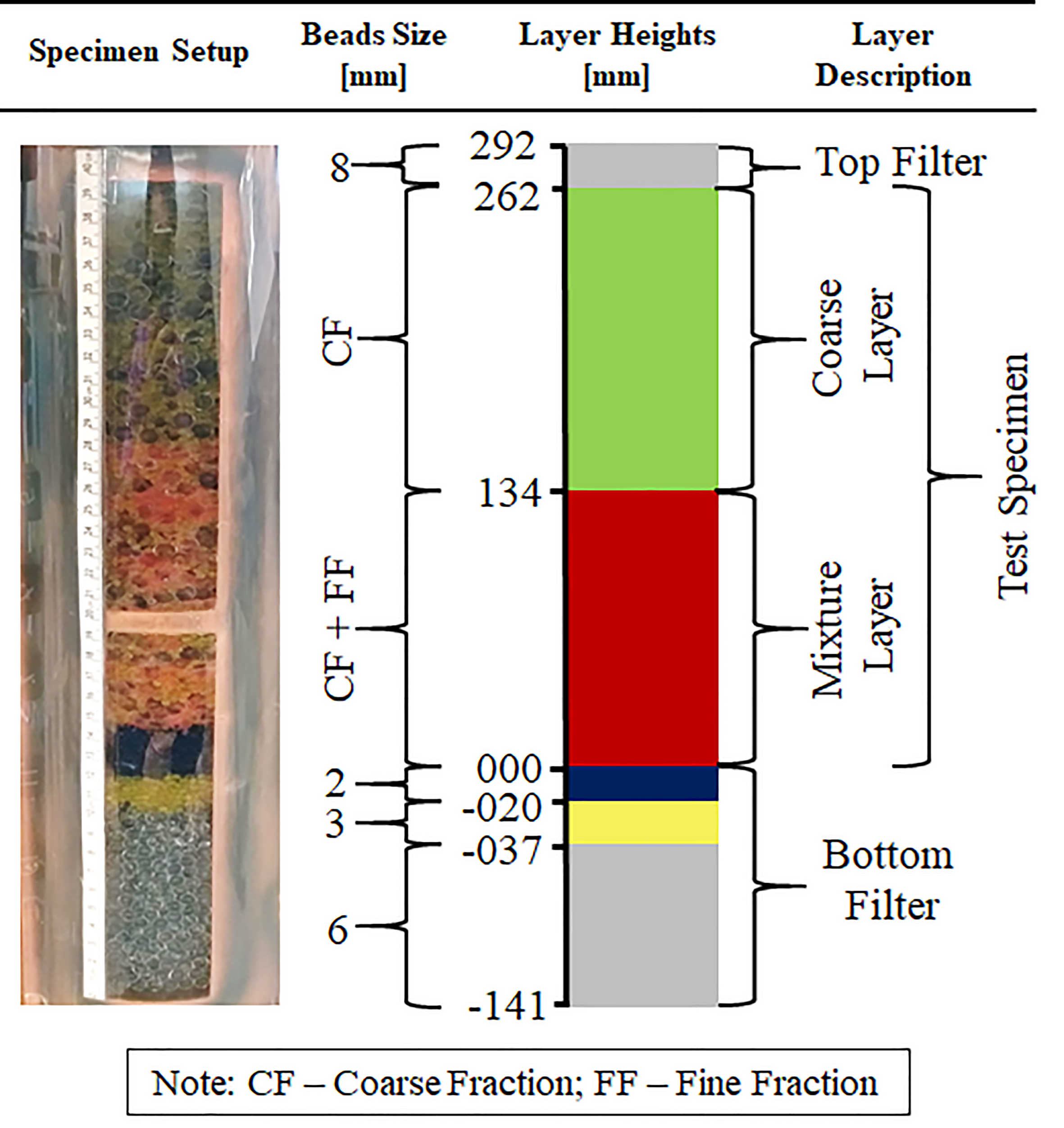}
     \captionof{figure}{Description of the test specimen}
     \label{fig:2}
 \end{Figure}
 
  \begin{Figure}
     \centering
     \captionsetup{type=figure,justification=centering}
     \includegraphics[height=6.0cm]{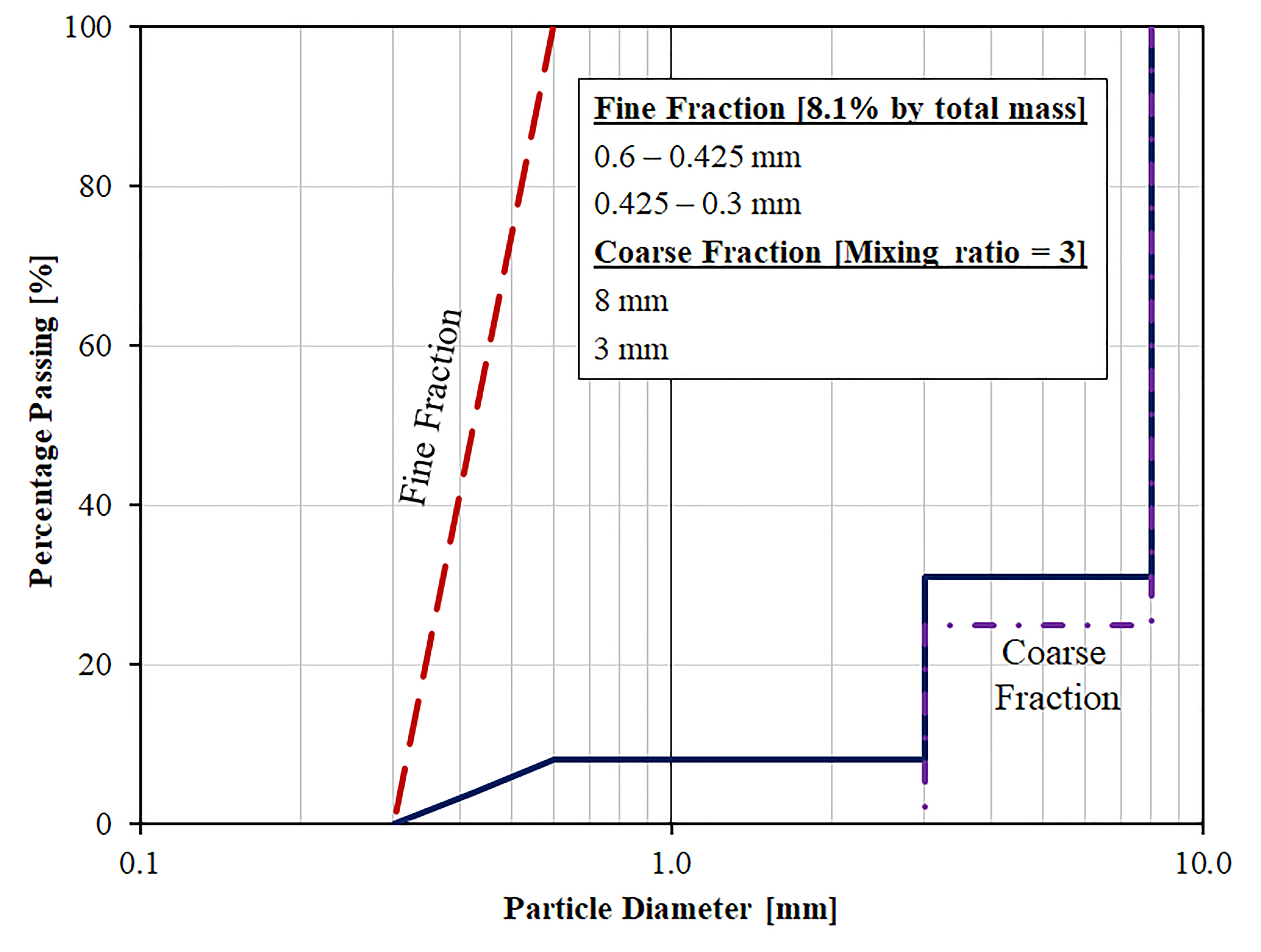}
     \captionof{figure}{Particle size distribution of the test specimen}
     \label{fig:3}
 \end{Figure}

\end{multicols}

\section{Methods to Compute the Permeability Profile}

\subsection*{Using average porosity and the original Kozeny-Carman Model}

The average permeability profile can be obtained from the original Kozeny-Carmen (K-C) model using the average porosity:
 \begin{equation}
     k_{average}=\frac{1}{C} \frac{n_{avg}^3}{\left(1-n_{avg}\right)^2} d_e^2
 \label{eq:1}
 \end{equation}
where $C$ is the shape factor, $n_{avg}$ is the average porosity of each layer which is estimated from the volume and dry weight of glass beads, and $d_e$ is the mean particle diameter. The shape factor accounts for grain shape effect and fluid flow heterogeneity, and for spherical particles, can be approximated as $C = 180$ \citep{zheng2017improved,carrier2003goodbye}. The mean particle diameter is obtained by discretising the PSD into equal-size bins and is given by:
 \begin{equation}
     d_e=\frac{1}{\Sigma_i\left(f_i/d_{li}^{0.405}d_{si}^{0.595}\right)}
 \label{eq:2}
 \end{equation}
where $f_i$ is the proportion of particles retained between the larger ($d_{li}$) and smaller ($d_{si}$) particle diameters of bin ‘$i$’.

\subsection*{Using point measurements of hydraulic head}

The point measurement of hydraulic heads using the pressure transducers (PT) along the height of the specimen can be used to calculate the local permeability between any two pressure transducers using equation (\ref{eq:3}):
\begin{equation}
    k_{local,PT}=\frac{Q}{i_{jk}A} \frac{\eta_w}{\gamma_w}
\label{eq:3}
\end{equation}
where $Q$ is the volumetric flow rate measured using the flowmeter, $i_{jk}=\Delta H/\Delta L$ is the gradient between two pressure transducers, where the subscript ‘$jk$’ represents the position of transducers, $\Delta H$ is the head drop and $\Delta L$ is the distance between the pressure transducers, $A$ is the cross-sectional area of the test specimen, $\eta_w$ is the dynamic viscosity of water and $\gamma_w$ is the water unit weight.

\subsection*{Using porosity measurements based on the STDR approach and a modified Kozeny--Carman Model}

The measured reflected TDR signal can be used to compute the porosity profile by applying a forward inversion model. The details on the inversion algorithm are provided in \citet{schlaeger2005fast}. The local permeability profile is then calculated from the local porosity profile using a modified Kozeny–-Carman (K--C) equation:
 \begin{equation}
     k_{local,TDR}^{K-C}=\frac{1}{C} \frac{n^3}{(1-n)^2}d_{eff}^2
 \label{eq:4}
 \end{equation}
where $n$ is the local porosity obtained from STDR measurements using the forward inversion model and $d_{eff}$ is the effective particle diameter for gap-graded soils. While equation (\ref{eq:2}) is the conventional method to obtain the effective particle diameter, it does not consider the influence of fine particle migration. For internally unstable soils under sufficiently large hydraulic gradients, the fine particles within the mixture layer may dislodge and migrate through the specimen. This leads to local changes in the effective particle diameter and porosity. To account for these local pore-scale changes to the soil structure, this study proposes the use of equation (\ref{eq:5}) to obtain the effective particle diameter:
 \begin{equation}
     d_{eff}^{-1}=\psi_f d_f^{-1} + \psi_c d_v^{-1}
 \label{eq:5}
 \end{equation}
where $d_f$ and $d_c$ are the mean particle diameters of the fine and coarse fractions, respectively, which are calculated from their respective PSDs using equation (\ref{eq:2}). $\psi_f$ and $\psi_c$ are the volume proportions of fine and coarse fractions, respectively. These can be calculated based on the porosity of fine and coarse fractions as detailed below.
The porosity of a soil is defined as:
 \begin{equation}
     n=\frac{V_t-V_s}{V_t}
 \label{eq:6}
 \end{equation}
where $V_t$ is the total volume and $V_s$ is the total solid volume. The porosity of the fine ($n_f$) and coarse ($n_c$) fraction is defined by:
\begin{equation}
    n_f=\frac{V_t-V_f}{V_t} \quad \mathrm{and} \quad n_c=\frac{V_t-V_c}{V_t}
\label{eq:7}
\end{equation}
where $V_f$ and $V_c$ is the solid volume of the fine and coarse fraction, respectively. From equations (\ref{eq:6}) and (\ref{eq:7}), the porosity of the soil can be expressed in terms of the porosity of fine and coarse fraction by:
\begin{equation}
    n=n_f+n_c-1
\label{eq:8}
\end{equation}
The volume proportion of fine and coarse fraction are given by:
\begin{equation}
    \psi_f=\frac{V_f}{V_f+V_c} \quad \mathrm{and} \quad \psi_c=\frac{V_c}{V_f+V_c}
\label{eq:9}
\end{equation}

Using equations (\ref{eq:6}) and (\ref{eq:7}), $\psi_f$ and $\psi_c$ can be expressed as a function of the various porosity terms:
\begin{equation}
    \psi_f=\frac{1-n_f}{1-n} \quad \mathrm{and} \quad \psi_c=\frac{1-n_c}{1-n}
\label{eq:10}    
\end{equation}
In equation (\ref{eq:10}), $n$ is inferred from STDR data using a forward inversion model, while $n_c$ is assumed to be constant, which is appropriate given the underfilled fabric of the test specimen. $n_f$ is derived based on the assumptions that the fine particles only redistribute within the specimen without leaving the sample and can be obtained from equation (\ref{eq:8}). This procedure enables the effective particle diameter to be determined as a basis for the calculation of the local permeability profile.

\subsection*{Using porosity measurements based on the STDR approach and a modified Katz--Thompson model}

In this method, the permeability profiles are computed from the porosity profiles using the modified Katz--Thompson (K--T) model:
 \begin{equation}
     k_{local,TDR}^{K-T}=\frac{\Lambda^2}{8F}
  \label{eq:11}
 \end{equation}
where $\Lambda$ is the interconnected pore radius which can be approximated by:
\begin{equation}
    \Lambda =\frac{d_{eff}}{2mF}
\label{eq:12} 
\end{equation}
where $F=n^{-m}$ is the electrical formation factor and $m$ [1.5 for spherical particles \citep{friedman2005soil}] is the cementation exponent \citep{glover2009cementation,bore2018new,mishra2021unified}.

\section{Results and Discussion}

\begin{figure}[ht]
     \centering
     \includegraphics[width=0.6\textwidth]{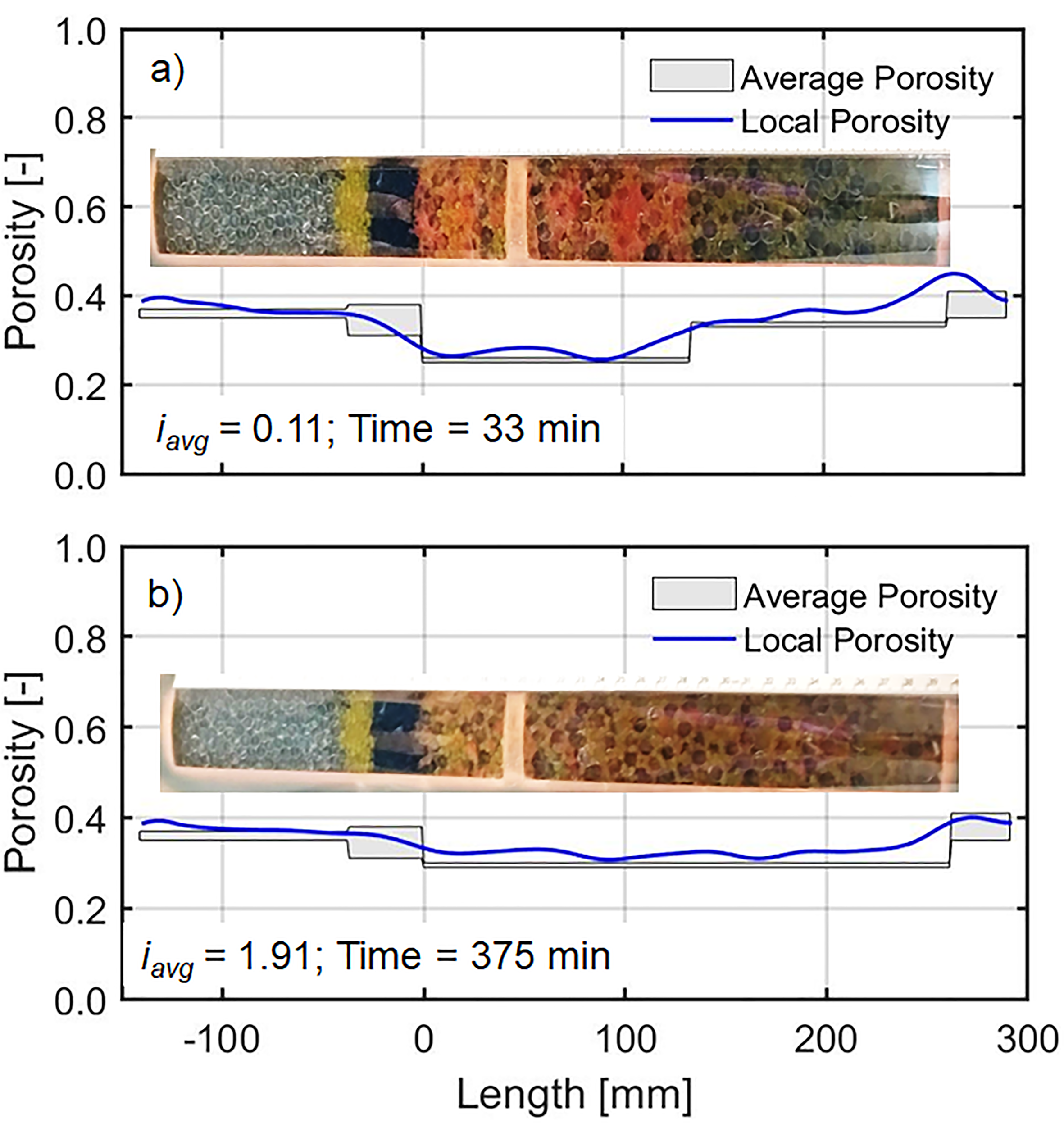}
     \caption{Comparison of the local porosity profile obtained from the STDR approach with average porosity profile a) at the initial state and b) at the final state. The average hydraulic gradient, $i_{avg}$, and the time at the initial and final state are shown in each figure}
     \label{fig:4}
 \end{figure}

The local permeability profile defined by $k_{local,TDR}^{K-C}$ and $k_{local,TDR}^{K-T}$ is dependent on the local porosity. The local porosity profile obtained from the STDR approach is compared with the average porosity obtained from the volume and dry weight of glass beads in each layer in Fig.~\ref{fig:4}. The local porosity profile shows a reasonable agreement with the average porosity, noting that the width of the average porosity profile reflects a 10\% tolerance on the layer heights. The smoothness of the local porosity profile at the layer transitions is a result of the rise time of the input TDR signal, where there is a trade-off between spatial resolution to identify layer transitions and minimising oscillations and noise in the porosity profile obtained using the forward inversion algorithm. By comparing the local porosity profile at the start of the test ($i_{avg} = 0.11$ and $T = 33$ min) in Fig.~\ref{fig:4}(a) with the end of the test ($i_{avg} = 1.91$ and $T = 375$ min) in Fig.~\ref{fig:4}(b), it can be seen that the porosity in the mixture layer (from length 0 to 134 mm) increased, and simultaneously, the porosity in the coarse layer (from length 134 to 262 mm) decreased. This is attributed to the migration of fine particles from the mixture layer to the coarse layer at a higher gradient ($i_{avg} = 1.91$). This can also be visually observed in the images of the test specimen shown in Fig.~\ref{fig:4}, where the red coloured particles indicate the fine fraction in the test specimen.

\begin{figure}[ht]
     \centering
     \includegraphics[width=0.70\textwidth]{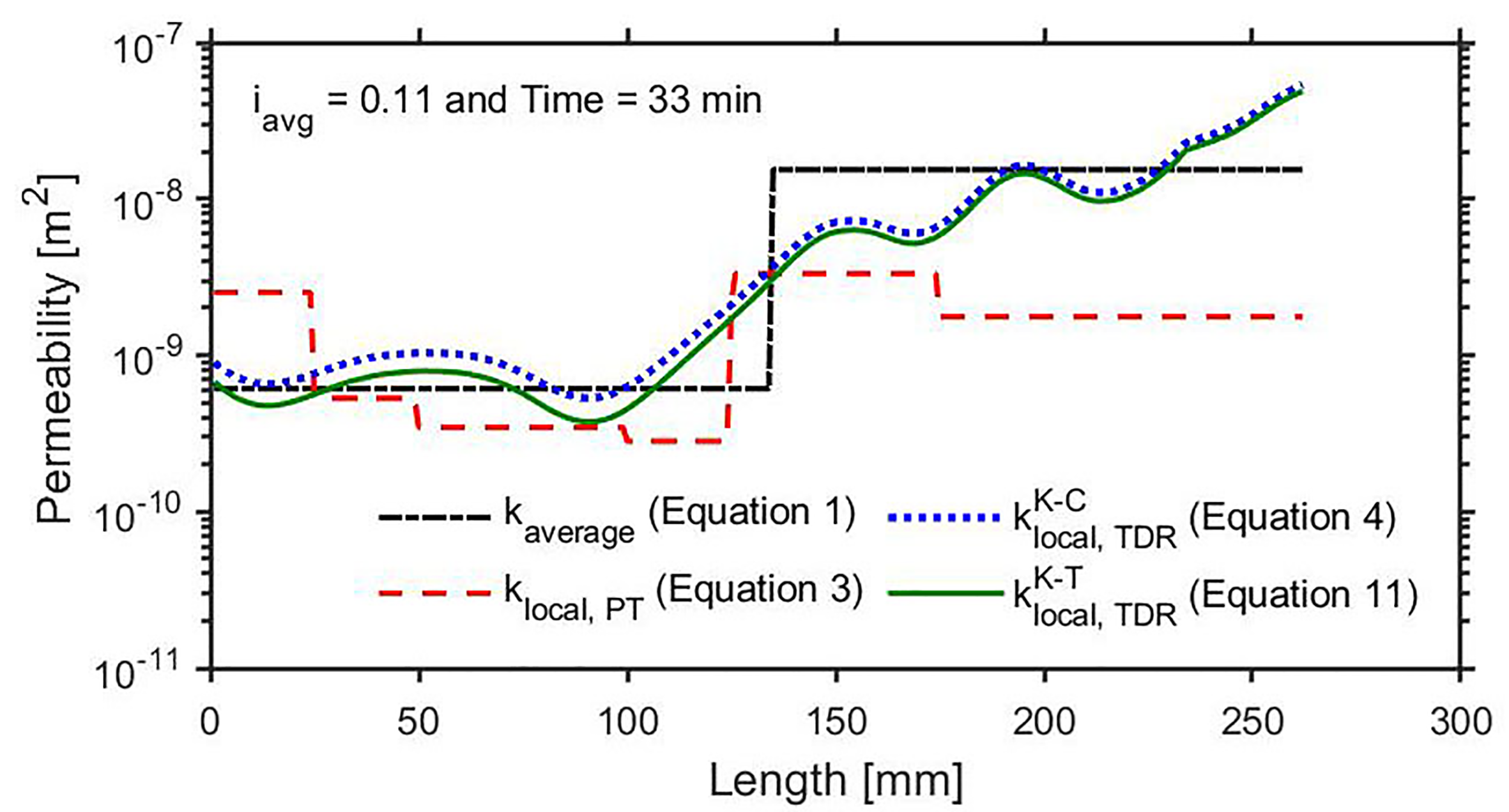}
     \caption{Permeability profile from four different approaches at the initial state of the experiment, including the average permeability, $k_{average}$ (equation \ref{eq:1}), local permeability based on pressure transducers, $k_{local,PT}$ (equation \ref{eq:3}), local permeability based on STDR measurements, $k_{local,TDR}^{K-C}$ (equation \ref{eq:4}) and $k_{local,TDR}^{K-T}$ (equation \ref{eq:11})}
     \label{fig:5}
 \end{figure}

 \begin{figure}[ht]
     \centering
     \includegraphics[width=0.7\textwidth]{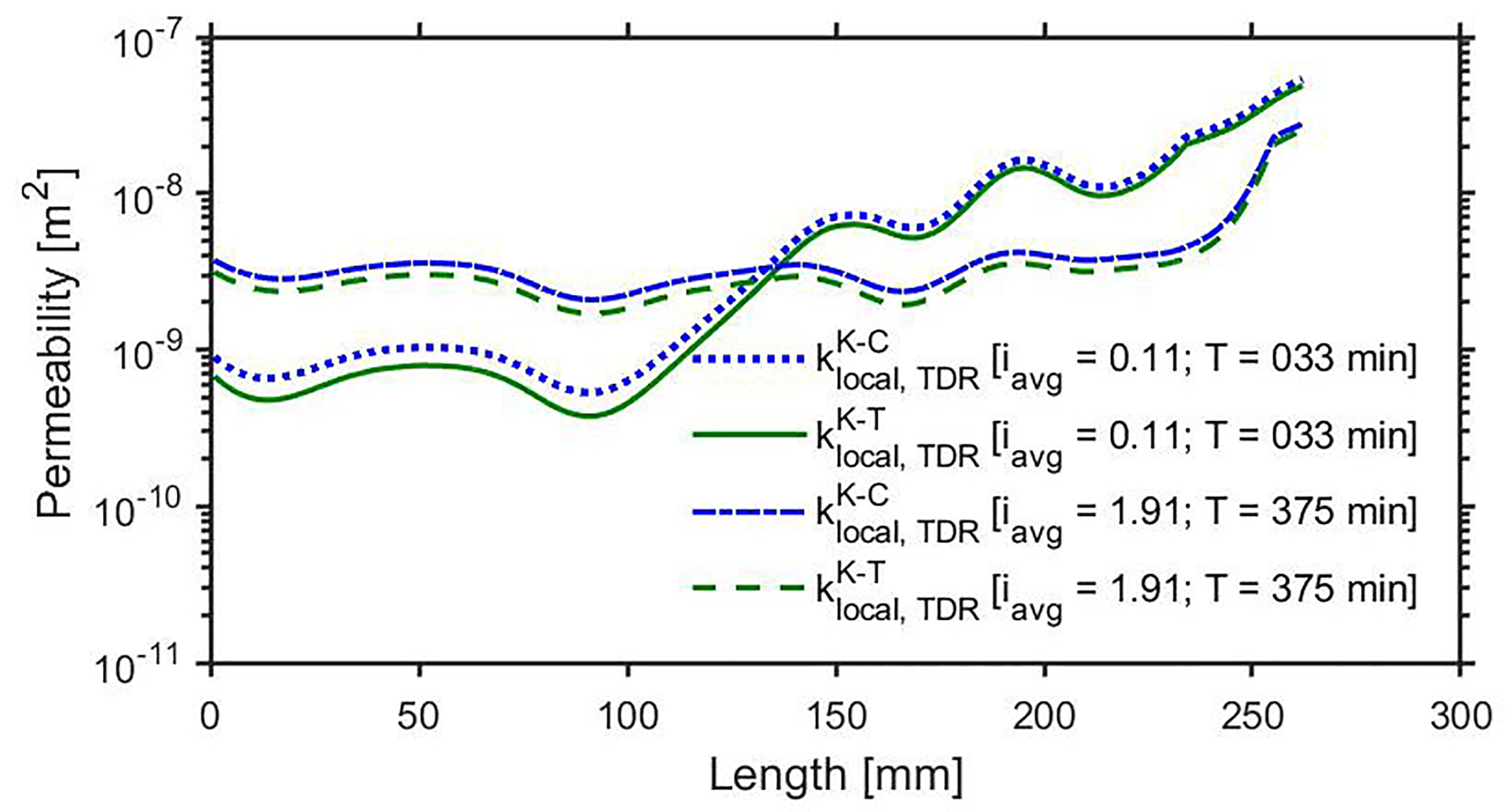}
     \caption{Comparison of local permeability profile from the modified K--C and K--T models at the initial and final states of the experiment}
     \label{fig:6}
 \end{figure}

Fig.~\ref{fig:5} illustrates the permeability profile of the test specimen computed from the four different methods at the initial state. A reasonable agreement is visually noted between all four methods. As expected, all four methods indicate that the permeability in the mixture layer is lower than that of the coarse layer. However, $k_{local,PT}$ in the coarse layer is lower than the other three methods ($k_{average}$, $k_{local,TDR}^{K-C}$ and $k_{local,TDR}^{K-T}$). This may be attributed to the flow condition in this layer, which is beyond the laminar range (Reynolds number, $R_e = 1.77 > 1.0$) even under a very small gradient of 0.11.

The average permeability, $k_{average}$, is constant over the mixture layer and the coarse layer, due to the assumption of homogenous layers with constant porosity and effective particle diameter. In contrast, all three measures of local permeability vary within each layer, as the local porosity may differ from the average porosity due to inherent pore-scale heterogeneity induced during sample preparation [Fig.~\ref{fig:4}(a)].

\begin{figure}[ht]
     \centering
     \includegraphics[width=0.7\textwidth]{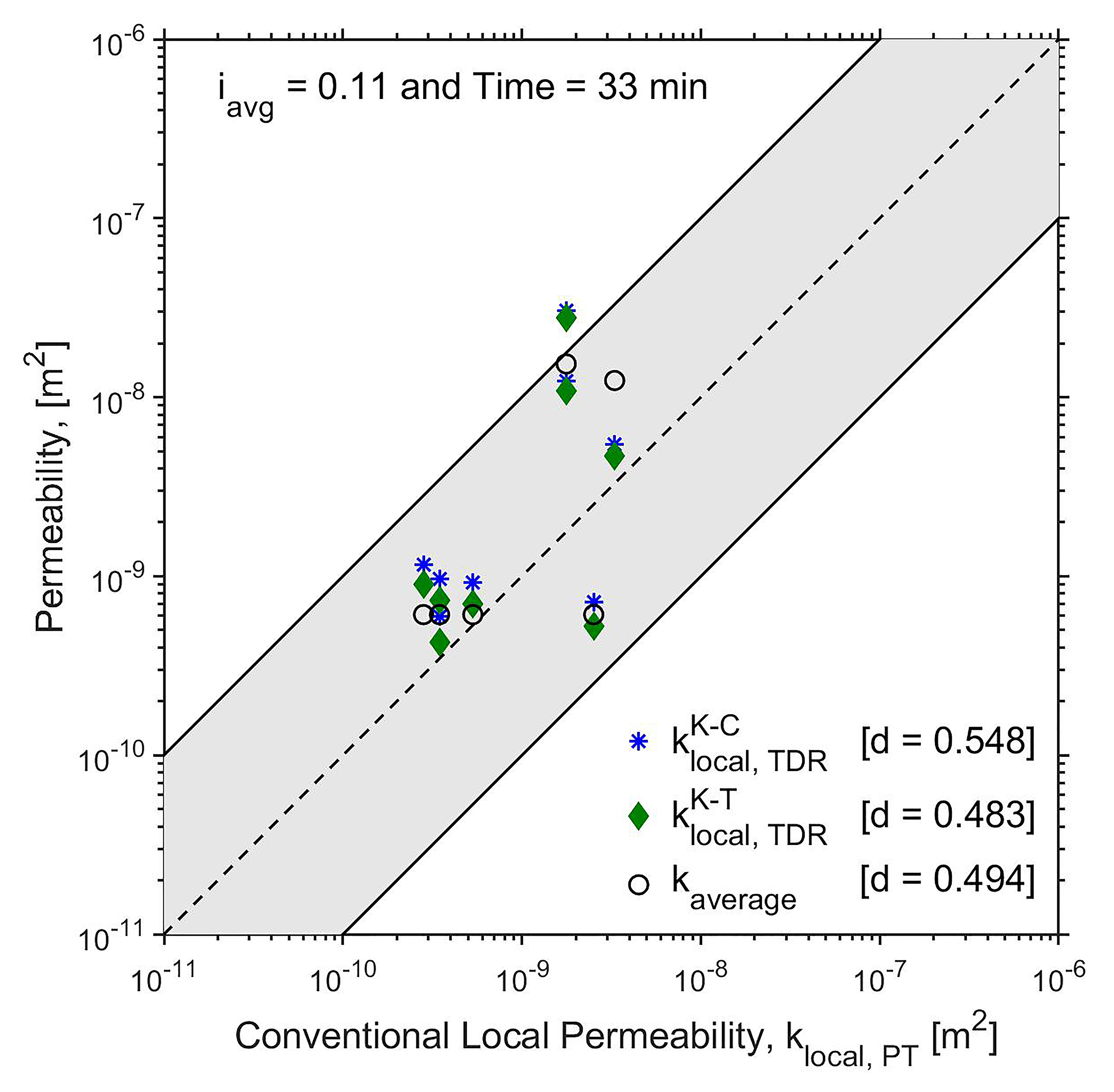}
     \caption{Comparison of conventional local permeability, $k_{local,PT}$ (equation \ref{eq:3}) against the average permeability, $k_{average}$ equation \ref{eq:1}) and the local permeability based on STDR, $k_{local,TDR}^{K-C}$  (equation \ref{eq:4}) and $k_{local,TDR}^{K-T}$ (equation \ref{eq:11}) at the initial state of the experiment. The average logarithmic deviation, $d$ is based on equation (\ref{eq:13})}
     \label{fig:7}
 \end{figure}

The local permeability profile using $k_{local,TDR}^{K-C}$ and $k_{local,TDR}^{K-T}$ are compared in Fig.~\ref{fig:6} at the initial and final states of the experiment. The permeability profile from the modified K--T model is slightly lower than that of the K--C model. As local permeability is conventionally obtained from pressure transducer measurements, Fig.~\ref{fig:7} compares $k_{local,PT}$  with the average permeability, $k_{average}$, and the local permeability from STDR measurements, $k_{local,TDR}^{K-C}$ and $k_{local,TDR}^{K-T}$. The local permeability profile for $k_{local,TDR}^{K-C}$ and $k_{local,TDR}^{K-T}$ is obtained at a higher spatial resolution compared to  $k_{local,PT}$, and to enable comparison with $k_{(local,PT)}$ in Fig.~\ref{fig:7}, $k_{local,TDR}^{K-C}$ and $k_{local,TDR}^{K-T}$ is averaged over the distance between the respective pressure transducers considered in obtaining $k_{local,PT}$. In Fig.~\ref{fig:7}, the 1:1 line is shown as a dashed line and the solid lines denote one order of magnitude of variation. To quantitatively compare the permeability values from the different methods, the mean absolute logarithmic deviation ($d$) between the permeabilities from the conventional ($c$) approach and proposed ($p$) methods is computed using equation (\ref{eq:13}):
 \begin{equation}
     d=\frac{\Sigma_{j=1}^N |log_{10}c_j - log_{10}p_j|}{N}
 \label{eq:13}
 \end{equation}
where $N$ denotes the total number of permeability values. $d=1.0$ indicates that the average deviation between the conventional local permeability, $k_{local,PT}$ and the proposed methods is one order of magnitude. Fig.~\ref{fig:7} shows that permeabilities are slightly overpredicted but fall within one order of magnitude ($d < 1.0$) from the conventional approach, which is acceptable for practical applications \citep{revil2015predicting,robinson2018permeability,weller2019permeability}.
 
\section{Conclusions}
This paper proposes semi-analytical methods to compute the local permeability profile in granular soils during hydraulic experiments. The proposed method uses a large co-axial cell permeameter in conjunction with spatial time-domain reflectometry to measure the spatial and temporal changes in porosity under applied hydraulic loading. The measured local porosity profile is used to calculate the local permeability using a modified Kozeny-Carman (K-C) and Katz-Thompson (K-T) equations by considering the influence of particle migration under sufficiently high hydraulic gradients. The local permeability profiles from these methods were shown to be comparable to the average permeability, which is assumed to be constant across a layer, and the conventional approach to obtaining local permeability from pressure transducers. These findings demonstrate the capabilities of spatial time-domain reflectometry in combination with suitable probe configurations to make physical observations on pore-scale heterogeneity.

\section*{Acknowledgements}
This work was funded by an Australian Research Council Future Fellowship awarded to A. Scheuermann (FT180100692) and an Australian Research Council Discovery Early Career Researcher Award accorded to T. Bore (DE180101441).

\bibliographystyle{ms}
\bibliography{ms}

\end{document}